\documentclass[12pt,a4wide]{article}
\usepackage{amsmath,amssymb,amsfonts}

\newcommand{\eqa}{\begin{eqnarray}}
\newcommand{\ena}{\end{eqnarray}}
\begin{document}
\begin{center}
{\large {\bf Birkhoff's Theorem for Quasi-Metric Gravity}}
\end{center}
\begin{center}
Dag {\O}stvang \\
{\em Department of Physics, Norwegian University of Science and Technology
(NTNU) \\
N-7491 Trondheim, Norway}
\end{center}
\begin{abstract}
Working within the quasi-metric framework (QMF), it is examined if the
gravitational field exterior to an isolated, spherically symmetric body is 
necessarily metrically static; or equivalently, whether or not Birkhoff's
theorem holds for quasi-metric gravity. It is found that it does; however
the proof is somewhat different from the general-relativistic case.
\\
\end{abstract}
\topmargin 0pt
\oddsidemargin 5mm
\renewcommand{\thefootnote}{\fnsymbol{footnote}}
\section{Introduction}
For General Relativity (GR), the validity of Birkhoff's theorem is well 
understood to be necessary both physically and mathematically. That is, just 
as Maxwell's equations forbid the existence of monopole electromagnetic waves, 
so do the Einstein field equations forbid the existence of monopole 
gravitational radiation [1], since according to Birkhoff's theorem, any 
time-dependent aspects of the space-time geometry interior to a spherically 
symmetric source cannot be propagated to the exterior gravitational field. 

Since the existence of monopole gravitational radiation is not desirable
for observational reasons, any potentially viable alternative theory of
gravity should also fulfil Birkhoff's theorem. In particular this applies to
quasi-metric gravity (the QMF is described in detail elsewhere [2, 3]). In 
this paper, equations relevant for the vacuum exterior to a spherically 
symmetric (in general metrically nonstatic) source are set up. It is then 
shown that no acceptable metrically nonstatic solutions exist, proving the 
validity of Birkhoff's theorem for quasi-metric gravity.
\section{Basic quasi-metric gravity}
The QMF has been described in detail elsewhere [2, 3]. Here we include only
the bare minimum of motivation and general formulae necessary to do the
calculations presented in later sections.

The basic motivation for introducing the QMF is the idea that the cosmic 
expansion should be described as a general phenomenon not depending on the 
causal structure associated with any pseudo-Riemannian manifold. This idea
would drastically reduce the enormous multitude of possibilities regarding 
cosmic genesis and evolution present in metric gravity and thereby increase the 
predictive power of the science of cosmology. And as we will see in what 
follows, certain properties intrinsic to quasi-metric space-time ensure that 
this alternative way of describing the cosmic expansion is mathematically 
consistent and fundamentally different from its counterpart in GR.

Briefly the geometrical basis of the QMF consists of a 5-dimensional 
differentiable manifold with topology ${\cal M}{\times}{\bf R}_1$, where 
${\cal M}={\cal S}{\times}{\bf R}_2$ is a Lorentzian space-time manifold, 
${\bf R}_1$ and ${\bf R}_2$ both denote the real line and ${\cal S}$ is a 
compact 3-dimensional manifold (without boundaries). That is, in addition to 
the usual time dimension and 3 space dimensions, there is an extra time 
dimension represented by {\em the global time function} $t$. (To ensure the 
uniqueness of $t$ (see below), the 3-dimensional manifold ${\cal S}$ is compact
by definition.) The reason for introducing this extra time dimension is that by 
definition, $t$ parameterizes any change in the space-time geometry that has 
to do with the cosmic expansion. By construction, the extra time dimension is 
degenerate to ensure that such changes will have nothing to to with causality. 
Mathematically, to fulfil this property, the manifold 
${\cal M}{\times}{\bf R}_1$ is equipped with two degenerate 5-dimensional 
metrics ${\bf {\bar g}}_t$ and ${\bf g}_t$. The metric ${\bf {\bar g}}_t$ is 
found from field equations as a solution, whereas the ``physical'' metric 
${\bf g}_t$ can be constructed from ${\bf {\bar g}}_t$ in a way described in 
refs. [2, 3].

The global time function is unique in the sense that it splits quasi-metric 
space-time into a unique set of 3-dimensional spatial hypersurfaces called 
{\em fundamental hypersurfaces (FHSs)} (where each FHS is represented by the 
3-manifold ${\cal S}$ for some epoch $t$). Observers always moving orthogonally
to the FHSs are called {\em fundamental observers (FOs)}. The topology of 
${\cal M}$ indicates that there also exists a unique ``preferred'' ordinary 
global time coordinate $x^0$. We use this fact to construct the 4-dimensional 
quasi-metric space-time manifold $\cal N$ by slicing the submanifold determined
by the equation $x^0=ct$ out of the 5-dimensional differentiable manifold. (It 
is essential that this slicing is unique since the two global time coordinates 
should be physically equivalent; the only reason to separate between them is 
that they are designed to parameterize fundamentally different physical 
phenomena.) Thus the 5-dimensional degenerate metric fields ${\bf {\bar g}}_t$ 
and ${\bf g}_t$ may be regarded as one-parameter families of Lorentzian 
4-metrics on $\cal N$. Note that there exists a set of particular coordinate 
systems especially well adapted to the geometrical structure of quasi-metric 
space-time, {\em the global time coordinate systems (GTCSs)}. A coordinate 
system is a GTCS iff the time coordinate $x^0$ is related to $t$ via $x^0=ct$ 
in ${\cal N}$.

Expressed in an isotropic GTCS, the most general form allowed for the family 
${\bf {\bar g}}_t$ is represented by the family of line elements valid on
the FHSs (this may be taken as a definition)
\eqa
{\overline {ds}}_t^2=[{\bar N}_{(t)}^s{\bar N}_{(t)s}-{\bar N}_t^2](dx^0)^2+
2{\frac{t}{t_0}}{\bar N}_{(t)i}dx^idx^0+
{\frac{t^2}{t_0^2}}{\bar N}_t^2{\tilde h}_{(t)ik}dx^idx^k.
\ena
Here, $t_0$ is some arbitrary reference epoch setting the scale of the spatial 
coordinates, ${\bar N}_t$ is the family of lapse functions of the FOs and 
${\frac{t_0}{t}}{\bar N}^k_{(t)}$ are the components of the shift vector family 
of the FOs in $({\cal N},{\bf {\bar g}}_t)$. Moreover,
${\bf {\bar h}}_t{\equiv}{\frac{t^2}{t_0^2}}{\bar N_t}^2{\bf {\tilde h}}_t$ is
the metric family on ${\cal S}$, i.e., the metric family intrinsic to the 
FHSs. Besides, 
${\bar N}_{(t)i}{\equiv}{\bar N}_t^2{\tilde h}_{(t)ik}{\bar N}_{(t)}^k$ (and thus
${\bar N}_{(t)}^j={\bar N}_t^{-2}{\tilde h}_{(t)}^{ij}{\bar N}_{(t)i}$).
Note that there are some prior-geometric restrictions on ${\bf {\tilde h}}_t$ 
(see below). Also note that, in order to define an affine connection being 
compatible with (the non-degenerate part of) ${\bf {\bar g}}_t$, we must have
that [2, 3]
\eqa
{\frac{\partial}{{\partial}t}}{\Big (}{\bar N}^i_{(t)}
{\bar N}^j_{(t)}{\tilde h}_{(t)ij}{\Big )}=0.
\ena 
One important interpretation of equation (1) is that gravitational quantities
should be ``formally'' variable when measured in atomic units. This formal
variability applies to all dimensionful gravitational quantities and is 
directly connected to the spatial scale factor ${\bar F}_t{\equiv}{\bar N}_tct$
of the FHSs [2, 3]. In particular, the formal variability applies to any 
potential gravitational coupling parameter $G_t$. It is convenient to transfer 
the formal variability of $G_t$ to mass (and charge, if any) so that all formal
variability is taken into account of in the {\em active stress energy tensor} 
${\bf T}_t$, which is the object that couples to space-time geometry via field 
equations. However, dimensional analysis yields that the gravitational coupling
must be non-universal, i.e., that the electromagnetic active stress-energy 
tensor ${\bf T}_t^{{\rm (EM)}}$ and the active stress-energy tensor for material 
particles ${\bf T}_t^{\rm (MA)}$ couple to space-time curvature via two different 
coupling parameters $G^{\rm B}$ and $G^{\rm S}$, respectively. This 
non-universality of the gravitational coupling  is required for consistency 
reasons and yields a modification of the right hand side of the gravitational 
field equations. (Said modification was missed in the original formulation of 
quasi-metric gravity.)

Moreover, due to the prior-geometric restriction on ${\bf {\tilde h}}_t$, a 
full coupling to space-time curvature of the active stress-energy tensor 
${\bf T}_t$ should not be expected to exist. But it turns out that a subset of 
the Einstein field equations (albeit modified) can be tailored to 
${\bf {\bar g}}_t$, so that {\em partial} couplings to space-time curvature of 
${\bf T}_t^{{\rm (EM)}}$ and ${\bf T}_t^{\rm (MA)}$ exist [2, 3]. The field 
equations then read (valid on the FHSs using a GTCS, and where a comma means 
taking a partial derivative)
\eqa
2{\bar R}_{(t){\bar {\perp}}{\bar {\perp}}}=
2(c^{-2}{\bar a}_{{\cal F}{\mid}i}^i+
c^{-4}{\bar a}_{{\cal F}i}{\bar a}_{\cal F}^i-
{\bar K}_{(t)ik}{\bar K}_{(t)}^{ik}+
{\cal L}_{{\bf {\bar n}}_t}{\bar K}_t) \nonumber \\
={\kappa}^{\rm B}(T^{{\rm (EM)}}_{(t){\bar {\perp}}{\bar {\perp}}}
+{\hat T}^{{\rm (EM)}i}_{(t)i})+
{\kappa}^{\rm S}(T^{{\rm (MA)}}_{(t){\bar {\perp}}{\bar {\perp}}}
+{\hat T}^{{\rm (MA)}i}_{(t)i}), \qquad c^{-2}{\bar a}_{{\cal F}i}{\equiv}
{\frac{{\bar N}_{t,i}}{{\bar N}_t}},
\ena
\eqa
{\bar R}_{(t)j{\bar {\perp}}}+{\Big (}{\frac{{\bar h}_{(t)}^{ik}}{{\bar N}_t}}
{\frac{\partial}{{\partial}x^0}}{\bar h}_{(t)ij}{\Big )}_{{\mid}k}-
{\Big (}{\frac{{\bar h}_{(t)}^{ik}}{{\bar N}_t}}
{\frac{\partial}{{\partial}x^0}}{\bar h}_{(t)ik}{\Big )},_j
={\bar K}_{(t)j{\mid}i}^i-{\bar K}_t,_j \nonumber  \\
+{\Big (}{\frac{{\bar h}_{(t)}^{ik}}{{\bar N}_t}}
{\frac{\partial}{{\partial}x^0}}{\bar h}_{(t)ij}{\Big )}_{{\mid}k}-
{\Big (}{\frac{{\bar h}_{(t)}^{ik}}{{\bar N}_t}}
{\frac{\partial}{{\partial}x^0}}{\bar h}_{(t)ik}{\Big )},_j
={\kappa}^{\rm B}T^{{\rm (EM)}}_{(t)j{\bar {\perp}}}
+{\kappa}^{\rm S}T^{\rm (MA)}_{(t)j{\bar {\perp}}}.
\ena
Here, ${\bf {\bar R}}_t$ is the Ricci tensor family corresponding to the metric
family ${\bf {\bar g}}_t$ and the symbol '${\bar {\perp}}$' denotes a scalar
product with $-{\bf {\bar n}}_t$, that is the negative unit normal vector field
family of the FHSs. Moreover, ${\cal L}_{{\bf {\bar n}}_t}$ denotes a projected Lie 
derivative in the direction normal to the FHSs, ${\bf {\bar K}}_t$ denotes the
extrinsic curvature tensor family (with trace ${\bar K}_t$) of the FHSs, a
``hat'' denotes an object projected into the FHSs and the symbol '${\mid}$'
denotes taking a spatial covariant derivative (compatible with 
${\bf {\bar h}}_t$). Finally, ${\kappa}^{\rm B}{\equiv}8{\pi}G^{\rm B}/c^4$ and
${\kappa}^{\rm S}{\equiv}8{\pi}G^{\rm S}/c^4$, where the values of $G^{\rm B}$ and 
$G^{\rm S}$ are by convention chosen as those measured in (hypothetical) local 
gravitational experiments in an empty universe at epoch $t_0$. Note that
the left hand side of equation (3) is similar to its counterpart in GR. On the 
other hand, the left hand side of equation (4) contains extra terms compared
to its counterpart in GR (${\bar h}_{(t)ij}$ are the components of the spatial 
metric family ${\bf {\bar h}}_t$ intrinsic to the FHSs). Said extra terms 
must be included for consistency reasons. 

In addition to the directly coupled field equations (3) and (4), we also have
a third field equation not involving any extra direct coupling to ${\bf T}_t$,
i.e., [2, 3]
\eqa
{\bar C}_{(t){\bar {\perp}}i{\bar {\perp}}j}={\tilde H}_{(t)ij}+
{\frac{1}{{\Xi}_0^2}}{\tilde h}_{(t)ij}, \quad {\Xi}_0{\equiv}ct_0, 
\ena
where ${\bf {\bar C}}_t$ is the Weyl tensor family in 
$({\cal N},{\bf {\bar g}}_t)$, and ${\bf {\tilde H}}_t$ is the spatial 
Einstein tensor family calculated from ${\bf {\tilde h}}_t$. (Note the last 
term on the right hand side of equation (5) yields a prior-geometric 
restriction on ${\bf {\tilde h}}_t$ since it implies that the corresponding
spatial Ricci scalar family ${\tilde P}_t={\frac{6}{{\Xi}_0^2}}$ is a 
constant.) Equation (5) may be written in the form [2, 3]
\eqa
{\frac{1}{{\bar N}_t}}{\cal L}_{{\bar N}_t{\bf {\bar n}}_t}{\bar K}_{(t)ij}
+{\bar K}_t{\bar K}_{(t)ij}-{\tilde H}_{(t)ij} \nonumber \\
={\frac{1}{3}}{\Big [}{\cal L}_{{\bf {\bar n}}_t}{\bar K}_t
+{\bar K}_t^2-2{\bar K}_{(t)ks}{\bar K}_{(t)}^{ks}+
{\frac{3}{(ct{\bar N}_t)^2}}{\Big ]}{\bar h}_{(t)ij}.
\ena
An explicit coordinate expression for ${\bf {\bar K}}_t$ may be calculated 
from equation (1). This expression reads (in a GTCS) [2, 3]
\eqa
{\bar K}_{(t)ij}={\frac{1}{2{\bar N}_t}}{\Big [}{\frac{t}{t_0}}
({\bar N}_{(t)i{\mid}j}+{\bar N}_{(t)j{\mid}i})-
{\frac{\partial}{{\partial}x^0}}{\bar h}_{(t)ij}{\Big ]},
\ena
\eqa
{\bar K}_t={\frac{t_0}{t}}{\frac{{\bar N}^i_{(t){\mid}i}}{{\bar N}_t}}
-{\frac{1}{2{\bar N}_t}}{\bar h}_{(t)}^{ij}{\frac{\partial}{{\partial}x^0}}
{\bar h}_{(t)ij},
\ena
and equations (7) and (8) have well-known counterparts in GR.
\section{Spherically symmetric exteriors in general}
We now set up the most general form for ${\bf {\bar g}}_t$ compatible with
the spherically symmetric condition. Introducing a spherically symmetric GTCS
${\{ }x^0,{\rho},{\theta},{\phi}{\} }$ where ${\rho}$ is an isotropic radial
coordinate, the spherically symmetric condition means that any shift vector
field must point in the ${\pm}{\rho}$-direction and that all unknown 
quantities at most depend on $t$, $x^0$ and ${\rho}$. Then equation (1) yields 
the family of line elements
\eqa
{\overline {ds}}_t^2=[{\bar N}_{(t)}^{\rho}{\bar N}_{(t){\rho}}-{\bar N}_t^2]
(dx^0)^2+2{\frac{t}{t_0}}{\bar N}_{(t){\rho}}d{\rho}dx^0+
{\frac{t^2}{t_0^2}}{\bar N}_t^2{\Big (}{\frac{{\tilde A}d{\rho}^2}
{1-{\frac{{\rho}^2}{{\Xi}_0^2}}}}+{\rho}^2d{\Omega}^2{\Big )}
\nonumber \\
={\bar B}{\Big [}-{\Big (}1-{\frac{
{\bar N}_{(t)}^{\rho}{\bar N}_{(t)}^{\rho}{\tilde A}}{
{1-{\frac{{\rho}^2}{{\Xi}_0^2}}}}}{\Big )}(dx^0)^2+2{\frac{t}{t_0}}
{\frac{{\bar N}_{(t)}^{\rho}{\tilde A}}{1-{\frac{{\rho}^2}{{\Xi}_0^2}}}}
d{\rho}dx^0+{\frac{t^2}{t_0^2}}{\Big (}{\frac{{\tilde A}d{\rho}^2}
{1-{\frac{{\rho}^2}{{\Xi}_0^2}}}}+{\rho}^2d{\Omega}^2{\Big )}{\Big ]},
\ena
where $d{\Omega}^2{\equiv}d{\theta}^2+{\sin}^2{\theta}d{\phi}^2$,
${\bar N}^{\rho}_{(t)}={\bar N}^{\rho}_{(t)}(x^0,{\rho},t)$,
${\tilde A}={\tilde A}(x^0,{\rho},t)$ and
${\bar B}={\bar B}(x^0,{\rho},t){\equiv}{\bar N}_t^2$. 
Note that the line element family (9) is by definition {\em metrically static} 
iff ${\tilde A}={\tilde A}({\rho})$, ${\bar B}={\bar B}({\rho})$ and 
${\bar N}_{(t)}^{\rho}{\equiv}0$.

The nonvanishing components of the extrinsic curvature tensor 
${\bf {\bar K}}_t$ become (from equations (7), (8) and (9))
\eqa
{\bar K}_{(t){\rho}}^{{\rho}}={\frac{t_0}{t{\sqrt{\bar B}}}}
{\Big [}{\bar N}^{\rho}_{(t),{\rho}}+{\bar N}^{\rho}_{(t)}{\Big (}
{\frac{{\tilde A}_{,{\rho}}}{2{\tilde A}}}+
{\frac{{\bar B}_{,{\rho}}}{2{\bar B}}}
+{\frac{{\rho}}{{\Xi}_0^2(1-{\frac{{\rho}^2}{{\Xi}_0^2}})}}
{\Big )}{\Big ]}-{\frac{{\tilde A}_{,0}}{2{\sqrt{\bar B}}{\tilde A}}}-
{\frac{{\bar B}_{,0}}{2{\bar B}^{3/2}}},
\ena
\eqa
{\bar K}_{(t){\theta}}^{{\theta}}={\bar K}_{(t){\phi}}^{{\phi}}=
{\frac{t_0}{t{\sqrt{\bar B}}}}{\bar N}_{(t)}^{\rho}{\Big (}
{\frac{{\bar B}_{,{\rho}}}{2{\bar B}}}+{\frac{1}{\rho}}{\Big )}-
{\frac{{\bar B}_{,0}}{2{\bar B}^{3/2}}},
\ena
\eqa
{\bar K}_t={\frac{t_0}{t{\sqrt{\bar B}}}}
{\Big [}{\bar N}^{\rho}_{(t),{\rho}}+{\bar N}^{\rho}_{(t)}{\Big (}
{\frac{{\tilde A}_{,{\rho}}}{2{\tilde A}}}+
{\frac{3{\bar B}_{,{\rho}}}{2{\bar B}}}+{\frac{2}{\rho}}
+{\frac{{\rho}}{{\Xi}_0^2(1-{\frac{{\rho}^2}{{\Xi}_0^2}})}}
{\Big )}{\Big ]}-{\frac{{\tilde A}_{,0}}{2{\sqrt{\bar B}}{\tilde A}}}-
{\frac{3{\bar B}_{,0}}{2{\bar B}^{3/2}}}.
\ena
Now the constraint equation (4) for spherically symmetric vacuum yields,
after some straightforward calculations, that
\eqa
{\bar N}_{(t)}^{\rho}{\Big [}{\frac{{\bar B}_{,{\rho}{\rho}}}{{\bar B}}}-
{\frac{3}{2}}{\Big (}{\frac{{\bar B}_{,{\rho}}}{{\bar B}}}{\Big )}^2-
{\Big (}{\frac{{\bar B}_{,{\rho}}}{2{\bar B}}}+{\frac{1}{\rho}}{\Big )}
{\Big (}{\frac{{\tilde A}_{,{\rho}}}{{\tilde A}}}
+{\frac{2{\rho}}{{\Xi}_0^2(1-{\frac{{\rho}^2}{{\Xi}_0^2}})}}{\Big )}{\Big ]}
\nonumber \\
+{\frac{t}{t_0}}{\Big [}{\frac{{\bar B}_{,0{\rho}}}{{\bar B}}}-
{\frac{3}{2}}{\frac{{\bar B}_{,0}{\bar B}_{,{\rho}}}{{\bar B}^2}}
-{\Big (}{\frac{{\bar B}_{,{\rho}}}{2{\bar B}}}+{\frac{1}{\rho}}{\Big )}
{\frac{{\tilde A}_{,0}}{{\tilde A}}}{\Big ]}=0.
\ena
Equation (13) is a (nonlinear) partial differential equation involving three 
unknown functions ${\tilde A}$, ${\bar B}$ and ${\bar N}_{(t)}^{\rho}$.
To fulfil Birkhoff's theorem, no vacuum solution exterior to an isolated, 
spherically symmetric source should exist for this equation, besides the
trivial metrically static solution ${\tilde A}={\tilde A}^{\rm ms}=1$,
${\bar B}={\bar B}^{\rm ms}({\rho})$, ${\bar N}_{(t)}^{\rho}=0$ (see below).
\section{Birkhoff's theorem}
The first step in proving Birkhoff's theorem in GR involves elimination of
the nonzero offdiagonal components of the metric. This can be done by 
performing a simple coordinate transformation to a new time coordinate (see,
e.g. [4]). Similarly, a new time coordinate ${x^0}'$ defined from the 
differentials
\eqa
d{x^0}'={\eta}({\rho},x^0,t){\Big [}1-{\frac{
{\bar N}_{(t)}^{\rho}{\bar N}_{(t)}^{\rho}{\tilde A}}{
{1-{\frac{{\rho}^2}{{\Xi}_0^2}}}}}dx^0-{\frac{t}{t_0}}
{\frac{{\bar N}_{(t)}^{\rho}{\tilde A}}{1-{\frac{{\rho}^2}{{\Xi}_0^2}}}}
d{\rho}{\Big ]},
\ena
where ${\eta}({\rho},x^0,t)$ is an integrating factor satisfying the
condition
\eqa
{\frac{\partial}{{\partial}{\rho}}}{\Big [}{\eta}{\Big (}1-{\frac{
{\bar N}_{(t)}^{\rho}{\bar N}_{(t)}^{\rho}{\tilde A}}{
{1-{\frac{{\rho}^2}{{\Xi}_0^2}}}}}{\Big )}{\Big ]}
=-{\frac{\partial}{{\partial}{x^0}}}{\Big [}{\eta}{\frac{t}{t_0}}
{\frac{{\bar N}_{(t)}^{\rho}{\tilde A}}{1-{\frac{{\rho}^2}{{\Xi}_0^2}}}}
{\Big ]},
\ena
eliminates the nonzero offdiagonal elements in equation (9). However, the
coordinate system ${\{ }{x^0}',{\rho},{\theta},{\phi}{\} }$ is not a GTCS
since we in general will have ${x^0}'{\neq}ct$ on the FHSs, i.e., the
hypersurfaces ${x^0}'=$constant cannot be identified with the FHSs. This would
be incovenient in the further analysis since quasi-metric spacetime directly
involves its foliation into the FHSs and not any other hypersurfaces. That is, 
since the basic formulae listed in section 2 will not in general be valid for 
a metric family foliated by hypersurfaces other than the FHSs, said elimination
would not be useful (but see [2, 3] for the possibility of having alternative 
foliations of quasi-metric space-time as a weak-field approximation for 
isolated systems in the limiting case ${\Xi}_0{\rightarrow}{\infty}$). For this
reason we will rather use equation (13) to prove that ${\bar N}_{(t)}^{\rho}$ 
must necessarily vanish for the gravitational field in vacuum outside an 
isolated spherically symmetric source.

To do that, we notice that equation (2) yields the condition
\eqa
{\frac{\partial}{{\partial}t}}{\Big (}{\bar N}^{\rho}_{(t)}
{\bar N}^{\rho}_{(t)}{\tilde A}{\Big )}=0.
\ena
But the $t$-dependence obtained from equation (13) is consistent with 
equation (16) only if ${\tilde A}$ is of the form 
${\tilde A}={\frac{t_0^2}{t^2}}{\tilde a}$, where ${\tilde a}$ does not depend
on $t$. But this form of ${\tilde A}$ is inconsistent with the general form (1)
of the metric family ${\bf {\bar g}}_t$. Thus we cannot have 
${\bar N}^{\rho}_{(t)}{\neq}0$, provided that the expressions in the square 
brackets of equation (13) do not vanish. On the other hand, if these 
expressions do vanish, i.e., if
\eqa
{\frac{{\bar B}_{,{\rho}{\rho}}}{{\bar B}}}-
{\frac{3}{2}}{\Big (}{\frac{{\bar B}_{,{\rho}}}{{\bar B}}}{\Big )}^2-
{\Big (}{\frac{{\bar B}_{,{\rho}}}{2{\bar B}}}+{\frac{1}{\rho}}{\Big )}
{\Big (}{\frac{{\tilde A}_{,{\rho}}}{{\tilde A}}}
+{\frac{2{\rho}}{{\Xi}_0^2(1-{\frac{{\rho}^2}{{\Xi}_0^2}})}}{\Big )}=0, \\
{\frac{{\bar B}_{,0{\rho}}}{{\bar B}}}-
{\frac{3}{2}}{\frac{{\bar B}_{,0}{\bar B}_{,{\rho}}}{{\bar B}^2}}
-{\Big (}{\frac{{\bar B}_{,{\rho}}}{2{\bar B}}}+{\frac{1}{\rho}}{\Big )}
{\frac{{\tilde A}_{,0}}{{\tilde A}}}=0,
\ena
one might still have that ${\bar N}^{\rho}_{(t)}{\neq}0$, however. But this
possibility only works if a limiting solution of equation (17) in the case of
no time dependence is the metrically static vacuum solution 
${\tilde A}^{\rm ms}=1$, ${\bar B}^{\rm ms}({\rho})$ found in [5] (see equation 
(27) below). Since said metrically static vacuum solution is not a solution of
equation (17) in the metrically static limit, the necessary correspondence does
not exist and we must necessarily have ${\bar N}^{\rho}_{(t)}{\equiv}0$.

To complete the proof of Birkhoff's theorem for the QMF, it remains to show 
that the metrically static solution found in [5] (see equation (27) below),
is the only possible vacuum solution exterior to a spherically symmetric body.
To achieve this, we will use equations (3) and (5) with 
${\bar N}^{\rho}_{(t)}{\equiv}0$.

Now we notice that since the Weyl tensor is conformally invariant, we have that
${\bar C}^{\alpha}_{(t){\beta}{\mu}{\nu}}={\tilde C}^{\alpha}_{(t){\beta}{\mu}{\nu}}$ and 
thus ${\bar C}_{(t){\bar {\perp}}i{\bar {\perp}}j}=
{\tilde C}_{(t){\tilde {\perp}}i{\tilde {\perp}}j}$, where ${\bf {\tilde C}}_t$ is the
Weyl tensor family calculated from the metric family ${\bf {\tilde g}}_t{\equiv}
{\bar N}_t^{-2}{\bf {\bar g}}_t$. The counterpart to equation (6) obtained from
equation (5) with ${\tilde C}_{(t){\tilde {\perp}}i{\tilde {\perp}}j}$ substituted for
${\bar C}_{(t){\bar {\perp}}i{\bar {\perp}}j}$ then reads
\eqa
{\cal L}_{{\bf {\tilde n}}_t}{\tilde K}_{(t)ij}
+{\tilde K}_t{\tilde K}_{(t)ij}-{\tilde H}_{(t)ij}
={\frac{1}{3}}{\Big [}
{\cal L}_{{\bf {\tilde n}}_t}{\tilde K}_t
+{\tilde K}_t^2-2{\tilde K}_{(t)ks}{\tilde K}_{(t)}^{ks}
+{\frac{3}{c^2t^2}}{\Big ]}{\frac{t^2}{t_0^2}}{\tilde h}_{(t)ij},
\ena
where ${\bf {\tilde K}}_t$ is the extrinsic curvature tensor family of the 
FHSs in the metric family ${\bf {\tilde g}}_t$. Now the 
${\rho}{\rho}$-component of equation (19) yields (with 
${\bar N}^{\rho}_{(t)}{\equiv}0$)
\eqa
{\frac{t^2}{t_0^2}}{\Big (}1-{\frac{{\rho}^2}{{\Xi}_0^2}}{\Big )}^{-1}
{\Big [}-{\frac{1}{3}}{\tilde A}_{,00}+{\frac{1}{6}}
{\frac{({\tilde A}_{,0})^2}{\tilde A}}{\Big ]}={\tilde H}_{(t){\rho}{\rho}}
+{\frac{1}{{\Xi}_0^2}}{\tilde h}_{(t){\rho}{\rho}}
={\frac{1}{{\rho}^2}}{\Big (}1-{\tilde A}{\Big )},
\ena
whereas twice the ${\theta}{\theta}$-component (or equivalently, twice
the ${\phi}{\phi}$-component) yields
\eqa
{\frac{t^2}{t_0^2}}{\Big [}{\frac{1}{3}}{\frac{{\tilde A}_{,00}}{\tilde A}}-
{\frac{1}{6}}{\Big (}{\frac{{\tilde A}_{,0}}{\tilde A}}{\Big )}^2{\Big ]}
{\rho}^2=2{\tilde H}_{(t){\theta}{\theta}}
+{\frac{2}{{\Xi}_0^2}}{\tilde h}_{(t){\theta}{\theta}}
=-{\rho}{\frac{{\tilde A}_{,{\rho}}}{({\tilde A})^2}}
(1-{\frac{{\rho}^2}{{\Xi}_0^2}})+2{\Big (}1-{\frac{1}{\tilde A}}{\Big )}
{\frac{{\rho}^2}{{\Xi}_0^2}}.
\ena
Moreover, combining equations (20) and (21) yields an equation which can be 
integrated to obtain an expression for ${\tilde A}$, i.e.,
\eqa
{\frac{{\tilde A}_{,{\rho}}}{{\tilde A}(1-{\tilde A})}}
={\frac{1}{\rho}}{\Big (}{\frac{1-3{\frac{{\rho}^2}{{\Xi}_0^2}}}
{1-{\frac{{\rho}^2}{{\Xi}_0^2}}}}{\Big )}, \qquad \Rightarrow \qquad
{\tilde A}={\Big (}1+{\frac{\tilde{\xi}}{\rho}}
(1-{\frac{{\rho}^2}{{\Xi}_0^2}})^{-1}{\Big )}^{-1},
\ena
where the function ${\tilde {\xi}}={\tilde {\xi}}(x^0,t)$ does not depend on 
$\rho$. Substituting the expression (22) for ${\tilde A}$ back into equation 
(21) then yields an ordinary differential equation for ${\tilde {\xi}}$, i.e.,
\eqa
{\tilde {\xi}}_{,00}-{\frac{3}{2}}{\frac{({\tilde {\xi}}_{,0})^2}{{\rho}(1+
{\frac{\tilde {\xi}}{\rho}}-{\frac{{\rho}^2}{{\Xi}_0^2}})}}=
3{\frac{t_0^2}{t^2}}{\frac{\tilde {\xi}}{{\rho}^2}}(1+
{\frac{\tilde {\xi}}{\rho}}-{\frac{{\rho}^2}{{\Xi}_0^2}}).
\ena
However, it is straightforward to see that this equation does not have any 
other solutions than the trivial solution ${\tilde {\xi}}=0$. This implies that 
that we must have that ${\tilde A}={\tilde A}^{\rm ms}=1$.

Next, with ${\tilde A}=1$ equation (18) yields
\eqa
{\frac{{\bar B}_{,0{\rho}}}{{\bar B}}}-
{\frac{3}{2}}{\frac{{\bar B}_{,0}{\bar B}_{,{\rho}}}{{\bar B}^2}}=0,
\quad {\Rightarrow} \quad
{\bar B}({\rho},x^0,t)={\frac{C(t)}{[f({\rho},t)+g(x^0,t)]^2}},
\ena
where the solution is found from MAPLE. Besides, equation (3) yields
(with ${\tilde A}=1$ and ${\bar N}^{\rho}_{(t)}{\equiv}0$)
\eqa
(1-{\frac{{\rho}^2}{{\Xi}_0^2}}){\frac{{\bar B}_{,{\rho}{\rho}}}{{\bar B}}}+
(2-3{\frac{{\rho}^2}{{\Xi}_0^2}}){\frac{{\bar B}_{,{\rho}}}{{\rho}{\bar B}}}
+3{\frac{t^2}{t_0^2}}{\Big [}{\Big (}{\frac{{\bar B}_{,0}}{{\bar B}}}{\Big )}^2
-{\frac{{\bar B}_{,00}}{{\bar B}}}{\Big ]}=0.
\ena
The general solution of equation (25) is on the form (from MAPLE)
\eqa
{\bar B}({\rho},x^0,t)=F({\rho},t)G(x^0,t).
\ena
However, this form is inconsistent with the form of ${\bar B}$ found in
equation (24) and the metrically static solution (27) below unless
${\bar B}={\bar B}({\rho})$. That is, no changes in the interior gravitational
field of a spherically symmetric source can propagate to the exterior vacuum.
Nor can ${\bar B}$ depend on $t$ since there is no such dependence for
a metrically static source. Thus the unique solution for the spherically
symmetric vacuum exterior to a spherically symmetric body with coordinate 
radius ${\rho}_{\rm s}$ can be found from equation (25) with 
${\bar B}={\bar B}({\rho})$, yielding the metrically static solution [5]
\eqa
{\bar B}^{\rm ms}({\rho})=1-{\frac{r_{{\rm s}0}}{\rho}}
{\sqrt{1-{\frac{{\rho}^2}{{\Xi}_0^2}}}}, 
\qquad
{\frac{r_{s0}}{{\sqrt{1+{\frac{r_{s0}^2}{{\Xi}_0^2}}}}}}<{\rho}<{\Xi}_0.
\ena
(Note that it is not meaningful to extend the solution (27) to beyond 
${\rho}={\Xi}_0$ since the transformation 
${\bf {\bar g}}_t{\rightarrow}{\bf g}_t$ becomes singular for 
${\rho}={\Xi}_0$ [5].) Here, 
$r_{{\rm s}0}{\equiv}{\frac{2M^{\rm (MA)}_{t_0}G^{\rm S}}{c^2}}+
{\frac{2M^{\rm (EM)}_{t_0}G^{\rm B}}{c^2}}$ is the quasi-metric counterpart to the
Schwarzschild radius at epoch $t_0$ and
\eqa
M_{t}^{\rm (MA)}=c^{-2}{\int}{\int}{\int}{\sqrt{{\bar B}}}{\Big [}
T_{(t){\bar {\perp}}{\bar {\perp}}}^{\rm (MA)}+{\hat T}^{{\rm (MA)}i}_{(t)i}
{\Big ]}d{\bar V}_t, \nonumber \\
M_{t}^{\rm (EM)}=c^{-2}{\int}{\int}{\int}{\sqrt{{\bar B}}}{\Big [}
T_{(t){\bar {\perp}}{\bar {\perp}}}^{\rm (EM)}+{\hat T}^{{\rm (EM)}i}_{(t)i}
{\Big ]}d{\bar V}_t,
\ena
are Komar masses corresponding to a metrically static source's content of 
material particles and electromagnetic fields, respectively [5]. If the source
is not metrically static the solution (27) is still valid, but not equation 
(28) for the active masses. Nevertheless, $r_{{\rm s}0}$ represents the active 
mass of the source at epoch $t_0$ as measured by distant orbiters. For some
later epoch $t_1>t_0$ the active mass measured will be represented by 
$r_{{\rm s}1}={\frac{t_1}{t_0}}r_{{\rm s}0}$, i.e., active mass increases linearly
with epoch independent of whether the source is metrically static or not.
Besides, performing a scaling of the radial coordinate
${\rho}{\rightarrow}{\rho}'={\frac{t_1}{t_0}}{\rho}$, the form of equation (9)
will be preserved with ${\bar N}^{{\rho}'}_{(t)}{\equiv}0$ and 
${\Xi}_0{\rightarrow}{\Xi}_1={\frac{t_1}{t_0}}{\Xi}_0$. This means that the 
secular increase of active mass does not depend on any form of communication 
between source and external gravitational field. Rather, the secular increase 
of active mass is just another facet of the global cosmic expansion as 
described within the QMF, i.e., a systematically changing relationship 
between dimensionful units as defined operationally from gravitational and
atomic systems, respectively. Thus there is no conflict between the secular
increase of active mass and the validity of Birkhoff's theorem for 
quasi-metric gravity.

Furthermore, just as for GR [1], in quasi-metric gravity Birkhoff's theorem 
holds for spherically symmetric electrovacuum outside an isolated charged 
source. This follows from the fact that $T^{\rm (EM)}_{(t)j{\bar {\perp}}}=0$ for 
this case (no radiation), so that all equations used showing the validity of 
the results ${\bar N}^{\rho}_{(t)}{\equiv}0$ and ${\tilde A}=1$ still hold. The 
only difference from the vacuum case is that equation (3) now has a source 
term, so that equation (25) gets a term ${\frac{r_{{\rm Q}0}^2}{{\rho}^4}}$ on 
the right hand side. That is, equation (25) changes to
\eqa
(1-{\frac{{\rho}^2}{{\Xi}_0^2}}){\frac{{\bar B}_{,{\rho}{\rho}}}{{\bar B}}}+
(2-3{\frac{{\rho}^2}{{\Xi}_0^2}}){\frac{{\bar B}_{,{\rho}}}{{\rho}{\bar B}}}
+3{\frac{t^2}{t_0^2}}{\Big [}{\Big (}{\frac{{\bar B}_{,0}}{{\bar B}}}{\Big )}^2
-{\frac{{\bar B}_{,00}}{{\bar B}}}{\Big ]}={\frac{r_{Q0}^2}{{\rho}^4}},
\quad r_{{\rm Q}0}{\equiv}{\frac{{\sqrt{2G^{\rm B}}}{\mid}Q{\mid}}{c^2}},
\ena
where $Q$ is the (passive) charge of the source [6]. But the solution of 
equation (29) is still of the general form shown in equation (26) and 
inconsistent with the solution form found in equation (24) if there is any
dependence on $x^0$. This again means that ${\bar B}={\bar B}({\rho})$ and it 
must be equal to the metrically static solution found in [6]. Note that in 
addition to the secular increase of active mass, the solution of equation (29) 
also implies a secular (linear) increase of {\em active charge} [6] 
contributing to ${\bf T}_t^{{\rm (EM)}}$.
\section{Conclusion}
In this paper it has been shown that Birkhoff's theorem is valid for 
quasi-metric gravity. It also holds for electrovacuum exterior to a charged, 
spherically symmetric, isolated source. There is no conflict between this 
result and the prediction that active mass as measured by distant test 
orbiters increases secularly with epoch [5] (see also active charge [6]); 
similar to the quasi-metric cosmic expansion, said prediction is a 
{\em global} phenomen not depending on any form of communicaton between the 
source and the external field.

Moreover, in quasi-metric gravity spherically symmetric exterior fields are 
not only isometric to the metrically static cases; in addition Birkhoff's 
theorem says that for said exterior fields, the FOs move exactly as for the 
metrically static cases.
\\ [4mm]
{\bf References} \\ [1mm]
{\bf [1]} C.W. Misner, K.S. Thorne and J.A. Wheeler, {\em Gravitation}, \\
{\hspace*{6.4mm}}W.H. Freeman {\&} Co. (1973). \\
{\bf [2]} D. {\O}stvang, {\em Grav. {\&} Cosmol.} {\bf 11}, 205
(2005) (gr-qc/0112025). \\
{\bf [3]} D. {\O}stvang, {\em Doctoral thesis}, (2001) (gr-qc/0111110). \\
{\bf [4]} S. Weinberg, {\em Gravitation and Cosmology}, J. Wiley {\&} Sons 
(1972). \\
{\bf [5]} D. {\O}stvang, {\em Grav. {\&} Cosmol.} {\bf 13}, 1 (2007)
(gr-qc/0201097). \\
{\bf [6]} D. {\O}stvang, {\em Grav. {\&} Cosmol.} {\bf 12} 262 (2006) 
(gr-qc/0303107).
\end{document}